# Fractal Nature of Solar Interior

Koushik Ghosh & Probhas Raychaudhuri

*Abstract:* Fractal method has been studied to understand the irregular and chaotic nature of any physical structure. Conventionally it is suggested that the solar interior is rigid in nature. Since solar neutrino flux is the indicator of the interior solar structure it is natural to study the solar neutrino flux source to find if the nuclear energy generation inside the sun is fractal in nature or not. At present there exist five solar neutrino experiments to detect neutrinos from the sun which can suggest which type of nuclear energy generation occurs inside the sun. Since we know that the solar atmosphere is irregular in nature many authors have studied this irregular nature by fractal analysis. In this regard we have studied solar neutrino flux data from 1) Homestake detector during the period from March, 1970 to April, 1994; 2) SAGE detector during the period from $1^{st}$ January, 1990 to $31^{st}$ December, 2000; 3) SAGE detector during the period from April, 1998 to December, 2001; 4) GALLEX detector during the period from May, 1991 to January, 1997; 5) GNO detector during the period from May, 1998 to December, 2001; 6) GALLEX-GNO detector (combined data) from May, 1991 to December, 2001; 7) average of the data from GNO and SAGE detectors during the period from May, 1998 to December, 2001; 8) 5-day-long samples from Super-Kamiokande-I detector during the period from June, 1996 to July,2001; 9) 10-day-long samples from Super-Kamiokande-I detector during the period from June,1996 to July,2001 and 10) 45-day-long samples from Super-Kamiokande-I detector during the period from June,1996 to July,2001 by fractal analysis and we have arrived at the conclusion that the solar neutrino flux data are fractal in nature.

*Index Terms:* **Solar neutrino flux data, fractal dimensions.**

## I. INTRODUCTION

Standard solar models (S.S.M.) are known to yield the solar internal structure with a overwhelming degree of success. The S.S.M. however cannot explain the solar activity cycle and solar cycle related variability. The variation of the sunspot numbers is one of the manifestation of the solar activity cycle. The observations have now revealed an 11 year cycle in the total solar irradiance with an amplitude of about 0.1% and is in phase with the solar magnetic activity cycle. The effective temperature and the radius of the sun also changes with the solar activity cycle. Solar oscillation frequencies are also known to change with time with the solar activity levels. Raychaudhuri [1,2] suggested a perturbed solar model with the hope to explain the solar cycle variations in the solar neutrino flux, total solar irradiance (luminosity), temperature and radius. Raychaudhuri [3] also explained the changes in oscillation frequencies. Raychaudhuri [4,5] suggested also five phases in all solar activities. S.S.M. cannot explain why the sunspot minimum range is shorter than sunspot maximum range. In general sunspot minimum range is about 4.6 years while the sunspot maximum range is about 6.4 years. Also it cannot explain why the first sunspot maximum to sunspot minimum takes about 6-6.4 years while for sunspot minimum to sunspot maximum it takes about 4-4.6 years. This can only be explained by the perturbed solar model suggested by Raychaudhuri [1,2].

Fractal method has been studied to understand the irregular and chaotic nature of graphical structure. Since fractals were introduced in physics, their applications promoted enormous progress in understanding phenomena that are most directly involved in formation of irregular structures. A broad class of clustering phenomena such as filtration, electrolysis and aggregation of colloids and aerosols have received a good deal of attention. Other phenomena that are not strictly clustering effects (i.e., dielectric breakdown, formation of a contact surface when two liquids are mixed etc.) can be advantageously treated using fractals. Describing natural objects by geometry is as old as science itself, traditionally this has involved the use of Euclidean lines, rectangles, cuboids, spheres and so on. But, nature is not restricted to Euclidean shapes. More than twenty years ago Benoit B. Mandelbrot observed that "clouds are not spheres, mountains are not cones, coastlines are not circles, bark is not smooth, nor does lightning travel in a straight line". Most of the natural objects we see around us are so complex in shape as to deserve being called geometrically chaotic. They appear impossible to describe mathematically and used to be regarded as the "monsters of mathematics". In 1975, Mandelbrot introduced the concept of fractal geometry to characterize these monsters quantitatively and to help us to appreciate their underlying regularity. The simplest way to construct a fractal is to repeat a given operation over and over again deterministically. The classical Cantor set is a simple text book example of such a fractal. It is created by dividing a line into $n$ equal pieces and removing $(n-m)$ of the parts created and repeating the process with $m$ remaining pieces ad infinitum. However, fractals that occur in nature occur through continuous kinetic or random processes. Having realized this simple law of nature, we can imagine selecting a line randomly at a given rate, and dividing it randomly, for example. We can further tune the model to determine how random this randomness is. Starting with an infinitely long line we obtain an infinite number of points whose separations are determined by the initial line and the degree of randomness with which intervals were selected. The properties of these points appear to be statistically self-similar and characterized by the fractal dimension, which is found to increase with the degree of increasing order and reaches its maximum value in the perfectly ordered pattern. It is now accepted that when the power spectrum of an irregular time series is expressed by a single power law $F^{-\alpha}$, the time series shows a property of a fractal curve. As the fractal length $L(k)$ of the time series is expressed as $L(k) \propto k^{-D}$ where $k$ is the time interval, the fractal dimension $D$ is expected to be closely related to the power law index $\alpha$. The relation between $\alpha$ and $D$ has been investigated by Higuchi [6] and it is given by $D = (5-\alpha)/2$.

We can determine the randomness of a time series by determining fractal dimensions and from this we can conclude whether a physical structure is chaotic in nature or not [7,8]. For any physical structure if the observed $D$ lies between 1 to 2 then we can conclude that the structure is chaotic and irregular in nature. For the ideal case of chaotic physical structure $D$ is 5/3, which describes inertial range turbulence in an incompressible fluid.

Conventionally it is suggested that the solar interior is rigid in nature. Since solar neutrino flux is the indicator of the interior solar structure it is natural to study the solar neutrino flux source to find if the nuclear energy generation inside the sun is fractal or not. In this context we have used Higuchi Method [9], Burlaga and Klein Method [10] and our method which is a modified form of Higuchi Method to analyze the solar neutrino flux data obtained from 1) Homestake detector during the period from March, 1970 to April, 1994[11]; 2) SAGE detector during the period from 1st January, 1990 to 31st December, 2000[12]; 3) SAGE detector during the period from April, 1998 to December, 2001[12]; 4) GALLEX detector during the period from May, 1991 to January, 1997[13]; 5) GNO detector during the period from May, 1998 to December, 2001[13]; 6) GALLEX-GNO detector (combined data) from May, 1991 to December, 2001[13]; 7) average of the data from GNO and SAGE detectors during the period from May, 1998 to December, 2001[12,13]; 8) 5-day-long samples from Super-Kamiokande-I detector during the period from June, 1996 to July,2001[14]; 9) 10-day-long samples from Super-Kamiokande-I detector during the period from June,1996 to July,2001[14] and 10) 45-day-long samples from Super-Kamiokande-I detector during the period from June,1996 to July,2001[14]. The analysis is carried in order to find the fractal dimensions of the solar neutrino flux data collected from different detectors as mentioned above, to see whether internal solar structure is chaotic and irregular in nature or not.

## II. CALCULATION OF FRACTAL DIMENSION

1. Higuchi Method:

Higuchi [9] developed a new method for calculating the fractal dimension of a given time series. Higuchi's method is as follows.

We take a finite set of time series taken at a regular interval:

$X(1), X(2), X(3),…,X(N)$.

From the given time series, we construct a new time series,

$\{X(m), X(m+k), X(m+2k), …, X(m + [(N − m) / k ] . k )\}$

where [ ] denotes the greatest integer function and both $k$ and $m$ ($m$ = 1, 2, 3, …, $k$) are integers, $m$ and $k$ indicate the initial time and the interval time respectively. Then $k$ sets of new time series are obtained. We define the length of the curve of the new time series as follows:

$$L_m(k) = \left\{ \left( \sum_{i=1}^{[(N-m)/k]} | X(m+ik) - X(m+(i-)k) | \right) \frac{N-1}{[(N-m)/k].k} \right\} / k \quad (1)$$

The length of the curve for the time interval $k$, $\langle L(k) \rangle$ is defined as the average value over $k$ sets of $L_m(k)$. If $\langle L(k) \rangle \propto k^{-D}$, we judge the curve is fractal with dimension $D$. We deduce fractal dimension $D$ from the slope of the best fitted line corresponding to the plot of $\log \langle L(k) \rangle$ against $\log k$.

2. Burlaga and Klein Method:

Burlaga and Klein [10] developed an interesting method to find out the fractal dimension of a given time series. The method is as follows:

We consider that $B(t)$ represents measurements of the observed magnitude, and the scale $\tau$ is determined by the averaging interval that we choose. The length of the curve $L(\tau)$, defined over some interval $0 \leq t \leq T_o$ (where $T_o = N\tau$ and $N$ is an integer) is

$$L(\tau) = \sum_{k=1}^{N} |B(t_k + \tau) - B(t_k)| \quad (2)$$

This length is a function of $\tau$ and for statistically self-affine curve, $L(\tau) = L_o \tau^{-S}$, where $L_o$ and $S$ are constants, and a plot of $\log L$ versus $\log \tau$ should be a straight line with a slope (−$S$). For practical purpose we deduce $S$ from the slope of the best-fitted line corresponding to the plot of $L(\tau)$ against $\tau$ on a doubly logarithmic scale, we introduce $D=S+1$, and this number $D$ is equal to the fractal dimension for statistically self-affine curves.

3. Modified Higuchi Method Developed by Authors:

We modified the Higuchi method by using weighted mean process to make data smooth. We write

$$L_m(k) = \left\{ \left( \sum_{i=1}^{[(N-m)/k]} |i.X(m+ik) - i.X(m+(i-1)k)| \right) \frac{(N-1)}{[(N-m)/k]([(N-m)/k]+1).(k/2)} \right\} / k \quad (3)$$

Next by similar process as described in Higuchi method we determine the fractal dimension $D$ from the slope of the best fitted line corresponding to the graph of the logarithm of length $\log \langle L(k) \rangle$ versus $\log k$.

### III. RESULTS AND CONCLUSION:

In the Higuchi method and modified Higuchi method we have calculated the fractal dimension $D$ directly whereas in the Burlaga and Klein method we first calculated $S$ and then we obtained the fractal dimension $D$ using the relation $D = S+1$. The obtained fractal dimensions $D$ for the solar neutrino flux data from different detectors are shown in the following tabular form :

| Experiments | Obtained Fractal Dimensions | | |
|---|---|---|---|
| | Higuchi Method | Burlaga & Klein Method | Modified Higuchi Method |
| Homestake (March, 1970 to April, 1994) | 1.995 | 2.040 | 1.989 |
| SAGE (1st January, 1990 to 31st December, 2000) | 1.949 | 1.935 | 1.920 |
| SAGE (April, 1998 to December, 2001) | 1.985 | 1.837 | 2.005 |
| GALLEX (May, 1991 to January, 1997) | 1.987 | 1.994 | 1.959 |
| GNO (May, 1998 to December, 2001) | 2.064 | 1.971 | 2.033 |
| GALLEX-GNO (Combined data) (May, 1991 to December, 2001) | 2.005 | 2.116 | 1.996 |
| Average of GNO and SAGE data (May, 1998 to December, 2001) | 2.012 | 1.941 | 1.984 |
| 5-day-long samples from Super-Kamiokande-I detector (June, 1996 to July, 2001) | 1.999 | 1.989 | 2.003 |
| 10-day-long samples from Super-Kamiokande-I detector (June, 1996 to July, 2001) | 1.990 | 1.994 | 1.993 |
| 45-day-long samples from Super-Kamiokande-I detector (June, 1996 to July, 2001) | 1.997 | 2.046 | 2.017 |

It is clear from the above table that the fractal dimensions of the solar neutrino flux data obtained from different detectors altogether lie between 1.837 and 2.116 from which we can confidently conclude that solar neutrino flux data obtained from different detectors are fractal in nature. Again, as the observed fractal dimensions lie around 2 i.e. the power law indices $\alpha$ will be around 1 we can say that it presumably represents a signal representing the generation of disturbances in the interior of the sun [15]. We investigated periodicities of solar neutrino flux data obtained from different detectors in another communication of ours[16] and we obtained that SAGE data(1st January 1990 to 31st December 2000) shows periodicity around 19 months,23.7-23.9 and 80.4-85.3 months; SAGE data(April 1998 to December 2001) gives periodicity around 1.5 months; GALLEX data(May 1991 to January 1997) shows periods around 1.6,2.5 and 18.6-18.9 months; GNO data(May 1998 to December 2001) gives periods around 1.4 and 3.4 months; combined GALLEX-GNO data(May 1991 to December 2001) shows periodicities around 1.7 and 18.5-19 months and average of the GNO and SAGE data(May 1998 to December 2001) gives periodicities around 1.8 and 2.2 months. We have also obtained the periodicities for 5-day-long samples from Super-Kamiokande-I detector (June 1996 to July 2001) at around 0.21-0.22,

0.67-0.77, 1.15-1.98, 6.72-6.95, 12.05-13.24, 22.48-24.02 months and for 10-day-long samples from Super-Kamiokande-I detector (June 1996 to July 2001) at around 0.39-0.45, 1.31-2.23, 5.20-5.32, 9.43-11.65 and 13.54-14.38 months[17]. These observations support the variable nature of solar neutrino and the observation of variable nature of solar neutrino would provide a significance to our understanding of solar internal dynamics and probably to the requirement of the modification of Standard Solar Model i.e. a perturbed solar model which is outlined by Raychaudhuri since 1971[1,2]. For the support of perturbed solar model we have demonstrated in this paper that solar neutrino flux data are fractal in nature. Hence, we can arrive at the conclusion that the solar interior appears to be perturbed and in this manner stochastic in nature.